\documentclass{article}
\usepackage{graphicx}
\makeindex
\begin{document}
\title{Efficient Monte Carlo Simulation Methods in Statistical Physics}
\author{Jian-Sheng Wang\\
Department of Computational Science,\\
National University of Singapore, Singapore 119260}
\date{15 March 2001}

\maketitle              
\begin{abstract}
The basic problem in equilibrium statistical mechanics is to compute
phase space average, in which Monte Carlo method plays a very
important role.  We begin with a review of nonlocal algorithms for
Markov chain Monte Carlo simulation in statistical physics.  We
discuss their advantages, applications, and some challenge problems
which are still awaiting for better solutions.  We discuss some of the
recent development in simulation where reweighting is used, such as
histogram methods and multicanonical method.  We then discuss the
transition matrix Monte Carlo method and associated algorithms.

The transition matrix method offers an efficient way to compute the
density of states.  Thus entropy and free energy, as well as the usual
thermodynamic averages, are obtained as functions of model parameter
(e.g. temperature) in a single run.

New sampling algorithms, such as the flat histogram algorithm and
equal-hit algorithm, offer sampling techniques which generate uniform
probability distribution for some chosen macroscopic variable.
\end{abstract}
\section{Statistical Physics}
Statistical mechanics \cite{reichl} is a theory developed at the end
of nineteenth century to deal with physical systems from an atomistic
point of view.  In principle the properties of bulk matter, which may
contain $10^{22}$ atoms, can be worked out from the motion of atoms
following the basic equations of Newtonian mechanics or quantum
mechanics.  However, such detailed information is not available or not
really necessary.  A probabilistic point of view, in the form of
statistical ensemble is taken.  The theory is extremely economical and
successful in dealing with {\sl equilibrium} problems.

There are a number of equivalent formulations of the theory.  But
statistical mechanics in a nutshell is the following concise formula
that connects statistical mechanics with thermodynamics.  First the
partition function is defined as
\begin{equation}
  Z = \sum_X \exp\left(-{E(X)\over kT} \right).
\end{equation}
where $X$ is a ``state'' of the system; the summation is carried over
all possible states.  $T$ is absolute temperature and $k$ is the
Boltzmann constant ($1.38 \times 10^{-23}$ Joule/Kelvin).  The free
energy of the system at temperature $T$ is given by
\begin{equation}
  F = - kT \ln Z. \label{free-energy}
\end{equation}
Free energy is a useful thermodynamic quantity in dealing with phase
transitions.  Other macroscopic observables are averages of the
corresponding microscopic quantities over the Boltzmann-Gibbs weight
$\exp\bigl(-E(X)/kT\bigr)$,
\begin{equation}
  \langle g(X) \rangle = 
           { 1\over Z} \sum_X g(X) \exp\bigl(-E(X)/(kT)\bigr).
\end{equation}
The following three quantities are perhaps the most important ones,
internal energy $U = \langle E \rangle$, heat capacity $C = dU/dT =
\bigl(\langle E^2 \rangle - \langle E \rangle^2\bigr)/(kT^2)$, 
and entropy $S = (U-F)/T = - k \langle \ln p(X) \rangle$, where the
average is over the Gibbs probability density, $p(X) =
\exp\bigl(-E(X)/kT\bigr)/Z$.

\section{Monte Carlo Method}
The essential task in statistical mechanics is to do the
multi-dimensional integrations (for continuous systems) or summations
(for discrete systems).  To fix the notation and language, we briefly
introduce the basics of Monte Carlo method
\cite{kalos,fishman,landau-binder}.  Consider the computation of a
statistical average,
\begin{equation}
  G = \int_\Omega g(X)\, p(X)\, dX = \langle g \rangle,
\end{equation}
where the probability density obeys $p(X) \ge 0$, and $\int_\Omega
p(X)\,dX = 1$.  Suppose that we can generate samples $X_i$ according
to the probability $p(X)$, then the integral can be estimated from an
arithmetic mean over the samples,
\begin{equation}
  G_M = { 1 \over M } \sum_{i=1}^M g(X_i).
  \label{estimator}
\end{equation}
The random variable $G_M$ has mean $G$ and standard deviation
$\sqrt{\tau \langle (g-G)^2 \rangle/M}$, $\tau$ being decorrelation
time.  Thus in the limit of large number of samples, the estimate
converges to the exact value.

The most general method to generate $X$ according to $p(X)$ is given
by a Markov chain \cite{norris} with a transition probability $P(X'|
X) = W(X \to X')$, satisfying the conditions stated below.  This is
the probability of generating a new state $X'$ given that the current
state is $X$.  Such process converges if
\begin{equation}
  W^n(X' \to X) > 0,\quad \mbox{for all $X'$ and $X$, and $n > n_0$}.
\end{equation}
The equilibrium distribution of the Markov chain satisfies
\begin{equation}
  p(X) = \sum_{X'} p(X') W(X' \to X).
\end{equation}
In constructing a Monte Carlo algorithm, it is convenient to consider
a much stronger condition, the detailed balance
\begin{equation}
  p(X') W(X' \to X) = p(X) W(X \to X').
\end{equation}

One of the most famous and widely used Monte Carlo algorithms is the
Metropolis importance sampling algorithm \cite{metropolis,hastings}.
It takes a simple choice of the transition matrix:
\begin{equation}
  W(X \to X') = S(X \to X') \min\left(1, {p(X') \over p(X)} \right)
\end{equation}
where $X \ne X'$, and $S$ is a conditional probability of choosing
$X'$ given that the current value is $X$, and it is symmetric, $S(X
\to X') = S(X' \to X)$. Usually $S(X \to X') = 0$ unless $X'$ is in
some ``neighborhood'' of $X$.  The diagonal term of $W$ is fixed
by the normalization condition $\sum_{X'} W(X \to X') = 1$.

\section{Cluster Algorithms}

In order to facilitate the discussion we first introduce the Ising
model.  Ising model is an interacting many-particle model for magnets.
A state consists of a collection of variables $\sigma_i$ taken on only
two possible values $+1$ and $-1$, signifying the spin up and spin
down states.  The spins are on a lattice.  The energy of the state
$\sigma = \{ \sigma_i \}$ is given by
\begin{equation}
  E(\sigma) = - J \sum_{\langle i,j\rangle} \sigma_i \sigma_j,
\end{equation}
where $J$ is a constant which fixes the energy scale, the summation is
over the nearest neighbor pairs.  When temperature $T$ is specified,
the states are distributed according to
\begin{equation}
  p(\sigma) = Z^{-1} \exp\left( - { E(\sigma) \over kT } \right).
\end{equation}

In a local Monte Carlo dynamics (Metropolis algorithm), one picks a
site $i$ at random, i.e., choosing a site with equal probability (this
specifies or realizes $S$).  Then, the energy increment if the spin is
flipped, $\sigma_i \to -\sigma_i$, is computed, which has the result
$\Delta E = 2 J \sigma_i \sum_j \sigma_j$.  The flip is accepted with
probability $\min\bigl(1, \exp(-\Delta E/kT) \bigr)$.  If the flip is
rejected, the move is also counted and the state remains unchanged.
One Monte Carlo step is defined as $N$ moves (trials) for a system of
$N$ spins, such that each spin is attempted to flip once on average.

The local algorithm of Metropolis type has some salient features: (1)
it is extremely general. Less assumption is made of the specific form
of probability distribution. (2) Each move involves $O(1)$ operations
and $O(1)$ degrees of freedom.  (3) The dynamics suffers from critical
slowing down.  The correlation time $\tau$ diverges as a critical
temperature is approached.  We shall elaborate more on this in the
following.

The statistical error, using estimator Eq.~(\ref{estimator}), is given
by
\begin{equation}
  \epsilon \approx \sigma_{G} \sqrt{ { \tau \over M}},
\end{equation}
where ${\rm Var}(G_1) = \sigma_G^2 =
\langle g^2 \rangle - \langle g \rangle^2$ is the
variance of the observable, $M$ is the number of Monte Carlo steps, and
$\tau$ is the decorrelation time.  We can take the point of view that the
above equation defines $\tau$, i.e.,
\begin{equation}
  \tau = \lim_{M \to \infty}{ M\, {\rm Var}(G_M) \over {\rm Var}(G_1) }. 
  \label{tau-def}
\end{equation}
Perhaps it is appropriate to call $\tau$ decorrelation time, since
$(\tau-1)/2$ is sometimes called correlation time in the literature.
The decorrelation time $\tau$ is the minimum number of Monte Carlo
steps needed to generate effectively independent, identically
distributed samples in the Markov chain.  The smallest possible value
for $\tau$ is 1, which represents independent sample by every step.
The usual integrated autocorrelation time \cite{sokal} differs from
our definition by a factor of 2, $\tau_{int} = \tau/2$.

The critical slowing down manifests itself by the fact that $\tau
\propto L^z$ at the critical temperature $T_c$ where a second order
phase transition occurs.  Here $L$ is the linear dimension of the
system ($N = L^d$ in $d$ dimensions).  For the local algorithms for
many models and in any dimensions, $z \approx 2$.  This suggests bad
convergence, specially for large systems.  At a first-order 
phase transition, where some thermodynamic variables change
discontinuously, the situation is even worse --- $\tau$ 
diverges exponentially with system sizes.

For the two-dimensional Ising model, a phase transition occurs
\cite{onsager} at $kT_c/J \approx 2.269$.  The magnetization is 
non-zero below this temperature and becomes zero above $T_c$. In
addition, in the limit of large system ($L \to \infty$), heat capacity
per spin and fluctuation of magnetization diverge.  These
intrinsic properties make computer simulation near critical point very
difficult.

Cluster algorithms \cite{swendsen-wang} overcome this difficulty
successfully.  For example, for the two-dimensional Ising model, the
dynamical critical exponent defined in $\tau \propto L^z$ is reduced
from 2.17 for the single-spin flip \cite{wang-gan} to much small value
for the cluster algorithm
\cite{swendsen-wang,heermann,coddington-PRL}.  It turns out that a
precise characterization of the Swendsen-Wang dynamical critical
exponent is very difficult, due to weak size dependence of the
decorrelation time.  Table~1 represents a recent extensive
calculation, based on the definition Eq.~(\ref{tau-def}) for the total
energy, rather than the usual method of extracting information from
time-dependent auto-correlation functions.  In this calculation, the
variance of the sum of $M$ consecutive energies
(c.f. Eq.~(\ref{estimator})) are computed explicitly.  Good
convergence to the limiting value is already achieved typically for $M
\approx 10^2$.  An $1/M$ extrapolation is used to get more accurate
estimates of the limit.  From this calculation, we find for the
two-dimensional Ising model, the convergence to asymptotics are slow.
It appears that the divergence is slightly faster than $\log L$.  If
we fit to power law $L^{z_{sw}}$, using two consecutive sizes of $L$
and $2L$, the exponent $z_{sw}$ decreases from 0.35 to 0.21.  It is
not clear whether $z_{sw}$ will convergence to a finite value or continue
decreasing to 0 in the limit of $L \to \infty$.

\begin{table}[t]
\center{
\setlength\tabcolsep{20pt}
\begin{tabular}{|r|l|l|}
\hline\hline
$L$   &    $\tau$      &       standard error \\
\hline
4   &  \hphantom{0}4.04575   &     0.00033  \\
8   &  \hphantom{0}5.17696   &     0.00032  \\
16  &  \hphantom{0}6.5165    &     0.0012   \\
32  &  \hphantom{0}8.0610    &     0.0018   \\
64  &  \hphantom{0}9.794     &	0.004    \\
128 &  11.723     &     0.012    \\
256 &  13.872     &     0.009    \\
512 &  16.29      &     0.04     \\
1024&  18.87      &     0.2      \\
\hline\hline
\end{tabular}
}
\caption[tab]{The decorrelation time $\tau$ of the Swendsen-Wang 
dynamics of the two-dimensional Isng model for different linear
lattice size $L$ at the critical temperature.}
\end{table}

The nonlocal cluster algorithm introduced by Swendsen and Wang for the
the Ising model (and more generally the Potts model) goes as follows:
(1) go over each nearest neighbor pair and create a bond with
probability
\begin{equation}
  1 - \exp\bigl( - 2 J \delta_{\sigma_i, \sigma_j}/(kT) \bigr).
\end{equation}
That is, if the two nearest neighbor spins are the same, a bond is
created between them with probability $1-\exp(-2J/kT)$; if spin values
are different, there will be no bond.  (2) Identify clusters as a set
of sites connected by zero or more bonds (i.e., connected component of
a graph).  Relabel each cluster with a fresh new value $+1$ or $-1$ at
random.

We note that each Monte Carlo step per spin still takes $O(1)$ in
computational cost.  The method is applicable to models containing
Ising symmetry, i.e., the energy is the same when $\sigma_i$ is
changed to $-\sigma_i$ globally.

The algorithm is based on a mapping of Ising model to a random cluster
model of percolation.  Specifically, we have
\cite{kasteleyn,coniglio,hu,edwards}
\begin{eqnarray}
  Z & = & \sum_{\{\sigma\}} \exp\Bigl(K \sum_{\langle i,j\rangle} 
       (\delta_{\sigma_i, \sigma_j} - 1)\Bigr) \\
    & = & \sum_{\{\sigma\}} \sum_{\{n\}} \prod_{\langle i,j\rangle} 
        \bigl[ p \delta_{\sigma_i, \sigma_j} \delta_{n_{ij},1} 
          + (1-p) \delta_{n_{ij},0} \bigr] \\
    & = & \sum_{\{n\}} p^b (1 - p)^{Nd - b} 2^{N_{\mathrm{c}}}, 
\end{eqnarray}
where $K = 2J/kT$, $p=1-\exp(-K)$, $\delta_{i,j}$ is the Kronecker
delta, and $b$ is the number of bonds, $d$ is the dimension of a
simple hypercubic lattice, and $N_{\mathrm{c}}$ is the number of
clusters.  $\sigma_i$ is spin on the site and $n_{ij} = 0,1$ is the
bond variable between the sites $i$ and $j$.  It is evident that the
moves in the Swendsen-Wang algorithm preserves configuration
probability of the augmented model containing both the spins and
bonds.

A single cluster variant due to Wolff \cite{wolff} is very easy to
program.  The following C code generates and flip one cluster:
\begin{verbatim}
      #define Z  4
      #define L  8
      double p = 0.7;
      int s[L*L];

      void flip(int i, int s0)
      {
         int j, nn[Z];
 
         s[i] = - s0;
         neighbor(i, nn);
         for(j = 0; j < Z; ++j)
            if(s0 == s[nn[j]] && drand48() < p)
               flip(nn[j], s0);
      }
\end{verbatim}
In this single cluster version, a site $i$ is selected at random.  The
value of the spin before flip is $s_0$.  It flips the spin and looks
for its neighbors.  If the value of the neighbor spins are the same as
$s_0$, with probability $p$ a neighbor site becomes part of the
cluster.  This is performed recursively.  It turns out that single
cluster is somewhat more efficient than the original Swendsen-Wang,
particularly in high dimensions.  For dimensions greater than or equal
to 4, Swendsen-Wang dynamics gives the dynamic critical exponent 
$z=1$, while Wolff single cluster is $z=0$ \cite{tamayo}.

It is easy to see why the single cluster algorithm of the above works.
Let us consider a general cluster flip algorithm with two bond
probabilities: $P_s$ will be the probability of connecting two
parallel spin sites; $P_d$ the probability of connecting anti-parallel
sites.  Consider the transition between two configurations $A$ and
$B$, characterized by flipping a cluster $C$.  A cluster is growing
from a site $i$ until the perimeter of the cluster is not connected to
the outside.  The transition probabilities can be written down as
\begin{eqnarray}
  W(A \to B) = P_{I} \prod_{\partial C(\uparrow\uparrow)}(1-P_s)
                     \prod_{\partial C(\uparrow\downarrow)} (1-P_d), \\
  W(B \to A) = P_{I} \prod_{\partial C(\uparrow\uparrow)}(1-P_d)
                     \prod_{\partial C(\uparrow\downarrow)} (1-P_s).
\end{eqnarray}
Note that the bond configuration probabilities are the same in the
interior of the clusters. The difference occurs at the boundary
$\partial C$, where parallel spins ($\uparrow\uparrow$) in $A$ becomes
anti-parallel spins ($\uparrow\downarrow$) in $B$, or vice versa.
Detailed balance requires that
\begin{equation}
  { W(A \to B) \over W(B \to A) } = { P(B) \over P(A)} = 
     (1-P_s)^{N_1 - N_2} (1 - P_d)^{N_2 - N_1} = 
     e^{-\Delta E \over kT},
\end{equation}
where $N_1$ is the number of parallel spins on boundary of $C$ in
configuration $A$; and $N_2$ is the number of anti-parallel spins on
boundary of $C$ in configuration $A$.  Since we have $\Delta E = E_B -
E_A = 2 (N_1 - N_2) J$, we obtain
\begin{equation}
  { 1 - P_s \over 1 - P_d } = \exp\left( - { 2 J \over kT } \right).
\end{equation}
Although the algorithm is valid for any $0 \leq P_d < 1$, it is most
efficient at $P_d = 0$, the Coniglio-Klein \cite{coniglio} bond
probability value.

A quite large number of statistical models can be treated with cluster
algorithms, with varied success. Excellent performance has been
obtained for Ising model, Potts models, and antiferromagnetic Potts
models \cite{wang-swendsen-kotecky}, XY model and general $O(n)$
models \cite{wolff}, $\phi^4$ field-theoretic model \cite{brower},
some regularly frustrated models \cite{kandel,coddington}, six-vertex
model \cite{evertz}, etc.  Cluster algorithms are proposed for hard
sphere fluid systems \cite{dress}, quantum systems
\cite{wiese,kawashima-loop,kawashima,beard}, microcanonical
ensembles \cite{creutz}, conserved order parameters
\cite{heringa,bikker}, etc.  Invaded cluster algorithm \cite{machta}
and other proposal \cite{tomita} are excellent methods for 
locating critical points.  The cluster algorithms are also
used in image processing \cite{besag,ferber}.

The cluster algorithms do not help much in temperature-driven
first-order phase transition -- the slow convergence has been shown
rigorously \cite{gor}.  Models with frustration, spin glass being the
archetype, do not have efficient cluster algorithms, although there
are attempts \cite{sw-spin-glass,liang,random-field} with limited
success.  Breakthrough in this area will have a major impact on the
simulation methods.

\section{Reweighting Methods}
In this and the following sections, we discuss a class of Monte Carlo
simulation approaches that aim at an efficient use of data collected,
and sampling methods that enhance rare events.

The computation of free energy, Eq.~(\ref{free-energy}), poses a
difficult problem for Monte Carlo method.  A traditional method is to
use thermodynamic integration, e.g.,
\begin{equation}
  { F(T_2) \over T_2 } - { F(T_1) \over T_1 } = 
            - \int_{T_1}^{T_2} { \langle E \rangle \over T^2 } dT,
\end{equation}
based on the relation $\langle E \rangle = - \partial \ln Z / \partial
\beta$, where $\beta = 1/(kT)$.

If we can estimate the density of states (the number of states with a
given energy $E$ for discrete energy models), then we can compute free
energy, as well as thermodynamic averages.  The result is obtained as
a function of temperature $T$, rather than a single datum point for a
specific value of $T$, as in standard Monte Carlo simulation.

This idea has been pursued over the last decade by Ferrenberg and
Swendsen \cite{ferrenberg,fs-multiple}, Berg et al
\cite{berg,berg-JSP,berg-review}, Lee \cite{lee}, Oliveira et al 
\cite{oliveira}, and Wang \cite{wang-euro}.  Consider the
following decomposition of summations over the states
\begin{equation}
  \sum_{\{\sigma\}} A(\sigma) e^{\left( - {E(\sigma) \over kT} \right)}
     = \sum_{E} e^{-{E \over kT}} \sum_{E(\sigma) = E} A(\sigma)
     = \sum_{E} e^{-{E \over kT}} n(E) \langle A(\sigma) \rangle_E,
\end{equation}
where $\langle A \rangle_E$ is the microcanonical ensemble average,
\begin{equation}
  \langle A \rangle_E = { 1\over n(E) } \sum_{E(\sigma) = E} A(\sigma).
\end{equation}
Since the state space is exponentially large ($2^N$ for the Ising
model with $N$ spins), and the range of $E$ is typically of order $N$,
if $n(E)$ can be computed accurately, the task is done.  The canonical
average of $A$ is related to the microcanonical average through
\begin{equation}
  \langle A \rangle_T = 
   { \sum_{E} \langle A \rangle_E \exp\left(-{E \over kT}\right) n(E) \over
    \sum_{E}  \exp\left(-{E \over kT}\right) n(E)},
\end{equation}
and free energy is computed as
\begin{equation}
  F = -kT \ln \sum_{E} \exp\left(- { E \over kT}\right) n(E).
\end{equation}

\subsection{Histogram method}
Ferrenberg and Swendsen \cite{ferrenberg} popularized a method which
in a sense is to compute the density of states (up to a multiplicative
constant) in a range close to a given simulation temperature.  This
method is generalized as multiple histogram method to combine
simulations at differential temperatures, to get the whole energy
range \cite{fs-multiple}.  We discuss here only the single histogram
method for its simplicity.

During a normal canonical simulation at fixed temperature $T^*$, we
collect the histogram of energy, $H(E)$, which is proportional to
probability distribution of energy,
\begin{equation}
  H(E) = c\, n(E) \exp\bigl(-E/(kT^*)\bigr).
\end{equation}
The constant $c$ is related to the partition function, $c = M/Z(T^*)$,
where $M$ is the total number of samples collected.  From the above
equation, we find $n(E) \propto H(E)\exp\bigl(E/(kT^*)\bigr)$.  With
this information, we can compute the free energy difference between
temperature $T^*$ and a nearby temperature $T$.  Similarly, moments of
energy can be computed after the simulation, through histogram
reweighting,
\begin{equation}
 \langle E^n \rangle_T = { \sum_E E^n H(E) \exp\bigl(-E/(kT) + E/(kT^*)\bigr)
  \over \sum_E H(E) \exp\bigl(-E/(kT) + E/(kT^*)\bigr)}.
\end{equation}

The range of $E$ that the histogram data can be collected at a fixed
temperature is limited by the energy distribution, which for the
canonical distribution away from critical point, is of order of
$\sqrt{N}$.  The whole range of energy $E$ is of order $N$.  This
limit the usefulness of single histogram method.

\subsection{Multicanonical Monte Carlo}
The multicanonical Monte Carlo method has been shown to be very
effective to overcome supercritical slowing down, reducing the
relaxation time from exponential divergence with respect to system
size to a power, at the first-order phase transitions \cite{berg}.
Multicanonical ensemble flattens out the energy distribution, so that
the computation of the density of states $n(E)$ can be done for all
values of $E$.  A multicanonical ensemble is defined to following the
probability density for the states as
\begin{equation}
	p(\sigma) = f\bigl(E(\sigma)\bigr) 
          \propto { 1 \over n\bigl(E(\sigma)\bigr) },
\end{equation}       
such that the energy histogram $H(E)$ ($\propto n(E) f(E)$) is a
constant.  From the histogram samples obtained by a simulation with
the weight of state at energy $E$ as $f(E)$, the density of state can
be computed \cite{lee} from $n(E) = H(E)/\bigl(f(E) c\bigr)$.  However, 
unlike canonical simulation where $f(E) = e^{-E/kT}$ is given, in a
multicanonical simulation, $f(E)$ is unknown to start with.

Berg proposed an iterative method to compute the weight in a
parametrized form $f(E) = \exp\bigl(-\beta(E) E + \alpha(E)\bigr)$,
starting with no information, $f_0(E) = {\rm const}$.  A new estimate
at iteration $n$ is then based on the results of all previous
iterations.  We refer to references \cite{berg-JSP,berg-review} for
details.

\section{Transition Matrix Monte Carlo and Flat Histogram Algorithm}

The flat histogram algorithm offers an efficient bootstrap to realize
the multicanonical ensemble, while transition matrix Monte Carlo
utilizes more data that can be collected in a simulation to improve
statistics.

\subsection{Transition matrix}

We start from the detailed balance equation for some given dynamics:
\begin{equation}
  p(\sigma) W(\sigma \to \sigma') = p(\sigma') W(\sigma' \to \sigma).
\end{equation}
By summation over the states $\sigma$ of fixed energy $E$, and
$\sigma'$ of fixed energy $E'$, and assuming that the probability of
the state is a function of energy only, $p(\sigma) \propto
f\bigl(E(\sigma)\bigr)$, we get
\begin{equation}
  n(E) f(E) T(E \to E') = n(E') f(E') T(E' \to E),
\label{T-balance}
\end{equation}
where the transition matrix in the space of energy is defined as
\begin{equation}
  T(E \to E') = { 1 \over n(E) } \sum_{E(\sigma) = E} \sum_{E(\sigma') = E'}
 W(\sigma \to \sigma').
\end{equation}
The matrix $T$ has a number of interesting properties: it is a
stochastic matrix in the sense of $T(E \to E') \ge 0$ and $\sum_{E'}
T( E \to E') = 1$; the stationary solution of $T$ is the energy
distribution $n(E)f(E)$; the dynamics associated with $T$ is
considerably faster than that of $W$ \cite{wang-tay-swendsen-PRL}.

We specialize to the case of single-spin-flip dynamics for the Ising
model.  The transition matrix $W$ for the spin states consists of a
product of two factors, the probability of choosing a spin to flip
$S(\sigma \to \sigma')$, and the flip rate $a(E \to E') = \min\bigl(1,
f(E')/f(E)\bigr)$.  We have $S(\sigma \to \sigma') = 0$ less the two
configurations $\sigma$ and $\sigma'$ differ by one spin, in this
case, the value of $S$ is $1/N$, where $N$ is the number of spins in
the system.  Using these results, we can rewrite the transition matrix
as
\begin{equation}
  T(E \to E') = { 1 \over N} \bigl\langle N(\sigma, E'-E) \bigr\rangle_E 
                a( E \to E'), \quad E \neq E'. \label{T-ssf}
\end{equation}
The diagonal elements are determined by normalization.  Substituting
Eq.(\ref{T-ssf}) into Eq.(\ref{T-balance}), using the relation between
$a(E \to E')$ and $f(E)$, we obtain
\begin{equation}
  n(E) \bigl\langle N(\sigma, E'-E) \bigr\rangle_E = 
  n(E') \bigl\langle N(\sigma', E-E') \bigr\rangle_{E'}.
  \label{broad-histo-eq}
\end{equation}
This is known as broad histogram equation
\cite{oliveira-europhys-b-2,berg-hansmann-europhys-b} which forms the
basis for the flat histogram algorithm presented below. Additionally,
this equation also gives us a way of computing the density of states
$n(E)$ by the quantity $\langle N(\sigma, E'-E)\rangle_E$ obtained 
from spin configurations generated from any distribution
$f\bigl(E(\sigma)\bigr)$.  The quantity
$N(\sigma, \Delta E)$ is the number of ways that the system goes to a
state with energy $E + \Delta E$, by a single spin flip from state
$\sigma$.  The angular brackets indicate a microcanonical average:
\begin{equation}
 \bigl\langle N(\sigma, \Delta E) \bigr\rangle_E = 
  { 1 \over n(E) } \sum_{E(\sigma) = E} N(\sigma, \Delta E).
\end{equation}

\subsection{Flat histogram algorithm}
The following algorithm \cite{wang-euro,wang-lee} generates flat
histogram in energy and realizes the multicanonical ensemble.
\begin{enumerate}
\item Pick a site at random.
\item Flip the spin with probability 
\begin{equation}
  a(E \to E') = \min\left(1, { \langle N(\sigma', -\Delta E) 
                                      \rangle_{E + \Delta E}
   \over \langle N(\sigma, \Delta E) \rangle_E } \right),
   \label{flat-histo-rate}
\end{equation}
where the current state $\sigma$ has energy $E$, the new state
$\sigma'$ has energy $E' = E + \Delta E$.
\item Accumulate statistics for $\langle N(\sigma, \Delta E) \rangle_E$.
\item Go to 1.
\end{enumerate}
We note by virtue of Eq.~(\ref{broad-histo-eq}), the flip rate is the
same as that in multicanonical simulation with a weight $1/n(E)$ and
Metropolis acceptance rate.  While in multicanonical sampling, the
weight is obtained through several simulations iteratively, the
quantities $\langle N(\sigma, \Delta E) \rangle_E$ is much easier to
obtain, through a single simulation.  This quantity serves a dual
purpose---it is used to construct a Monte Carlo algorithm (used as
input), and at the same time, it is used to compute the density of
states (output of the simulation).  Clearly, this is circular unless
approximation is made.  We have considered replacing the exact
microcanonical average by an accumulative average, over the history of
simulation generated so far, i.e.,
\begin{equation}
  \bigl\langle N(\sigma, \Delta E) \bigr\rangle_E \approx
{ 1\over H(E) } \sum_{i=1}^M \delta_{E(\sigma^i), E} N(\sigma^i,\Delta E),
\end{equation}
where $\{\sigma^i,\; i = 1, 2, \cdots\}$ is the sequence of states
generated with the algorithm given above; $H(E)$ is the number of
samples accumulated at the energy bin $E$.  In case the data for
computing the flip rate is not available, we simply accept the move to
explore the new state.  A more rigorous way of doing simulation is to
iterate the simulation with constant flip rate.  For example, after
the first simulation, we compute a first estimate to the density of
states. In a second simulation, we perform multicanonical simulation
\'a la Lee \cite{lee}.  The data collected in the second run for $\langle
N(\sigma, \Delta E) \rangle_E$ will be unbiased.  It is found that
even with a single simulation, the results converge to the exact
values (for $\langle N(\sigma, \Delta E) \rangle_E$ and $n(E)$) for
sufficiently long runs, even though a rigorous mathematical proof of
the convergence is lacking.

Wang and Landau \cite{wang-landau} proposed recently a new algorithm
that works directly with the density of states $n(E)$.  The simulation
proceeds with the flip rate $\min\bigl(1, n(E)/n(E')\bigr)$, but 
the value of the density of states is updated after every move by
$n(E) \leftarrow n(E) f$ and letting $f \to 1$ for convergence.
Excellent results were obtained.  A careful comparison with flat
histogram method is needed.

\subsection{N-fold way (rejection-free moves)}
In Metropolis algorithm, moves are sometimes rejected.  This rejection
is important for realizing the correct stationary distribution.  In
1975 Bortz, Kalos, and Lebowitz \cite{bortz} proposed a rejection-free
algorithm.  It is still based on Metropolis flip rate, but the waiting
due to rejection is taking into account by considering all possible
moves.  The Bortz-Kalos-Lebowitz N-fold way algorithm for the Ising
model goes as follows:
\begin{enumerate}
\item Compute the acceptance probability for one attempt of a move
\begin{equation}
  A = \sum_{\Delta E} { N(\sigma, \Delta E) \over N} a( E \to E + \Delta E).
\end{equation}
\item Pick an energy change $\Delta E$ according to probability
\begin{equation}
  p_{\Delta E} = 
    { N(\sigma, \Delta E) \over A\, N} a( E \to E + \Delta E).
\end{equation}
flip a site belonging to $\Delta E$ with probability 1.  The site is
choosing from the $N(\sigma, \Delta E)$ sites with equal probability.
\item One N-fold-way move is equivalent to $1/A$ moves in the original
dynamics.  Thus thermodynamic averages are weighted by $1/A$, i.e.,
$\langle g \rangle = \sum_{t} (g_t/A) \big/ \sum_{t} 1/A$, where summation
is over every move $t$. 
\end{enumerate}
In order to implement step 2 efficiently, additional data structure is
needed so that picking a spin in a given class characterized by
$\Delta E$ is done in $O(1)$ in computer time.

\subsection{Equal-hit algorithm}
Combining N-fold way and flat histogram algorithm is easy, since the
important quantity $N(\sigma, \Delta E)$ is already computed in flat
histogram algorithm.  The flip rate $a$ is given by formula
(\ref{flat-histo-rate}).  In the flat histogram algorithm, the
probability that the energy of the system is $E$ is a constant, i.e.,
\begin{eqnarray}
 H(E)& = & \langle \delta_{E(\sigma), E} \rangle \propto n(E) f(E) 
               = \mbox{const} \nonumber \\
     & = & \left\langle \delta_{E(\sigma), E} 
                       { 1 \over A(\sigma)} \right\rangle_N \Big/
        \left\langle { 1 \over A(\sigma) } \right\rangle_N.
\end{eqnarray}
The averages in the second line of the above equation refer to samples
generated in an N-fold way simulation.

In equal-hit algorithm (ensemble) \cite{swendsen-int}, we require that
the number of ``fresh configurations'' generated at each energy is a
constant.  More precisely equal-hit ensemble is {\sl defined} by
\begin{equation}
  u(E) = \langle \delta_{E(\sigma), E} \rangle_N = \mbox{const}.
\end{equation}
One possible choice of the flip rate is
\begin{equation}
  a(E \to E') = \min\left(1, { \langle {1 \over A} \rangle_{E',N} 
                \langle N(\sigma', E - E') \rangle_{E'} \over 
               \langle {1 \over A } \rangle_{E, N} 
                \langle N(\sigma, E'-E) \rangle_E } \right),
\end{equation}
where 
\begin{equation}
  \left\langle {1 \over A } \right\rangle_{E, N} = 
    { \langle \delta_{E(\sigma), E} { 1 \over A(\sigma) } \rangle_N \over
      \langle \delta_{E(\sigma), E} \rangle_N }
\end{equation} 
is the inverse total acceptance rate $1/A$ arithmetic averaged over the
N-fold way samples at energy $E$.

The histogram $H(E)$ generated in the equal-hit algorithm depends on
the precise dynamics (the rate $a$) used.  Since there are many
possible choices of the rate, such ``equal-hit ensemble'' is not
unique.

\subsection{Determination of the density of states}
While Eq.~(\ref{broad-histo-eq}) gives us a way of obtaining the
density of states, there are more equations than unknowns. We consider
two optimization methods. The first method is based on the transition
matrix itself.  We define $T_{E, \Delta E} = \langle N(\sigma, \Delta
E) \rangle_E/N$.  Symbols with hat being unknown, and $T_{E, \Delta
E}$ the Monte Carlo estimate, consider
\begin{equation}
  {\rm Minimize}\quad 
  \sum_{E, \Delta E} \sigma^{-2}_{E, \Delta E} 
      \bigl( \hat T_{E, \Delta E} - T_{E, \Delta E} \bigr)^2,
\end{equation}
subject to $0 \leq \hat T_{E, \Delta E} \leq 1$, $\sum_{\Delta E} \hat
T_{E, \Delta E} = 1$, and $\hat T_{E,1} \hat T_{E+1,1} \hat T_{E+2,-2}
= \hat T_{E,2} \hat T_{E+2,-1} \hat T_{E+1,-1}$.  The last constraint
needs more explanation.  We assume that the energy level is equally
spaced (as in the Ising model). Consider three energy levels, $E$,
$E+1$, $E+2$.  If we write down three equations of type
(\ref{broad-histo-eq}), for transitions from $E$ to $E+1$, $E+1$ to
$E+2$, and $E+2$ to $E$, we can cancel the density of states by
multiplying the three equations together.  This leaves the last
equation above, and it is known as TTT identity.  It can be shown that
multiple T identities (four or more) are not independent, and they
need not put in as constraints.  For Ising model there is also one
additional symmetry constraint, $T_{E,i} = T_{-E,-i}$.

When the solution for $T$ is found, we can use any of the energy
detailed balance equation to find density of states $n(E)$.  The TTT
identity guarantees that the answer is unique whichever detailed
balance equation is used.

The second method is based on optimization directly with variable
$n(E)$, actually $S(E) = \ln n(E)$, by
\begin{equation}
 {\rm minimize} \quad 
 \sum_{E, \Delta E} \sigma^{-2}_{E, \Delta E} 
   \left( \hat S(E+\Delta E) - \hat S(E) - 
          \ln { T_{E,\Delta E} \over T_{E+\Delta E, -\Delta E}} \right)^2,
\end{equation}
subject to, for the Ising model, $\sum_{E} n(E) = 2^N$, where $N$ is
the total number of spins in the system.  In addition, we can put in 
the known fact that the ground states are doubly degenerate, $n(0) = 2$.

\begin{figure}[t]
\includegraphics[width=.95\textwidth]{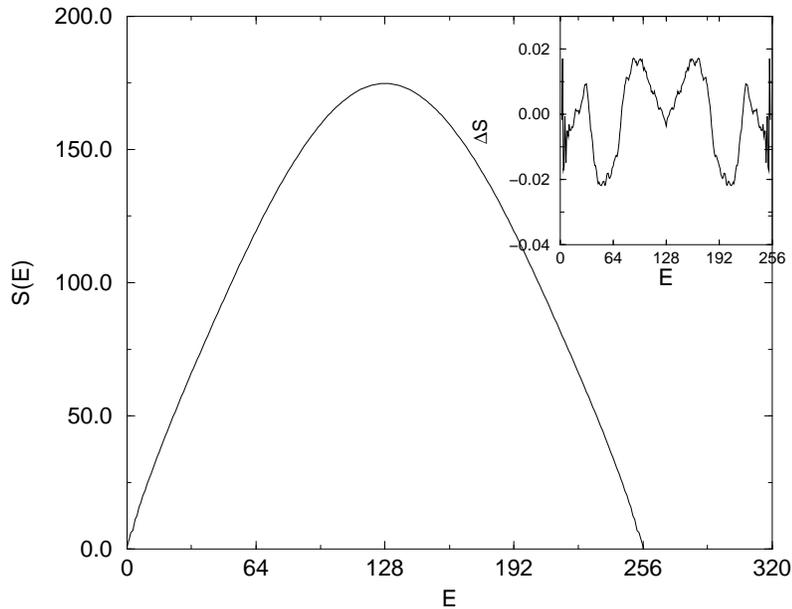}
\caption[fig2]{Logarithmic density of states, $S(E)= \ln n(E)$, 
calculated by transition matrix Monte Carlo method for a $16 \times
16$ two-dimensional Ising model with a total $1.1 \times 10^5$ Monte
Carlo steps and the first $10^4$ discarded, using the flat-histogram
algorithm and N-fold way. The insert is error in $S(E)$ with respect
to exact results.}
\end{figure}

Figure~1 shows one of the simulation results of the density of states,
using the second method.  The errors by comparison with exact known
values \cite{beale} are presented in the insert of the figure.  The density of
states $n(E)$ is determined to an accuracy of better than 2 percents
in a matter of few minutes of computer time.

The flat-histogram dynamics is used to study spin glasses \cite{zhan}.
The dynamic characteristics is quite similar to multi-canonical method
of Berg.  The study of lattice polymer and protein folding is under
way.  For related ideas and approaches, see
refs.\cite{munoz,lima,kastner}.

\section{Conclusion}

The phenomenon of critical slowing down can be effectively dealt with
by cluster algorithms for a large class of statistical mechanics
models.  We reported new and very accurate results for the
decorrelation times of the Swendsen-Wang dynamics.  Large size
asymptotic behavior is analyzed.  For super-critical slowing down
occurring in first-order phase transitions, multicanonical ensemble
simulation and flat-histogram or equal-hit algorithms are very
effective.  Since the latter algorithms and associated transition
matrix method is efficient in computing the density of states, this
method can also be useful for general counting problems by Monte Carlo
method.

\section*{Acknowledgements}
The author thanks Prof. R. H. Swendsen for many of the work discussed
here.  He also thank Z. F. Zhan, T. K. Tay, and L. W. Lee for
collaborations.  This work is supported by a NUS research grant
R151-000-009-112 and Singapore-MIT Alliance.

%
%


\begin{thebibliography}{99}
%
\bibitem{reichl}
Reichl, L. E.: A Modern Course in Statistical Physics, Wiley-Interscience,
(1998).

\bibitem{kalos}
Kalos, M. H., Whitlock, P. A.: Monte Carlo Methods, Vol I: Basics,
John Wiley \& Sons (1986).

\bibitem{fishman} Fishman, G. S.: Monte Carlo, concepts, algorithms,
and applications, Springer, (1996). 

\bibitem{landau-binder} Landau, D. P., Binder, K.:
A Guide to Monte Carlo Simulations in Statistical Physics, 
Cambridge Univ Press, (2000).

\bibitem{norris} 
Norris, J. R.: Markov Chains, Cambridge Univ Press, (1997).

\bibitem{metropolis}
Metropolis, N., Rosenbluth, A. W., Rosenbluth, M. N.,
Teller, A. H., Teller, E.: J. Chem. Phys. {\bf 21} (1953) 1087.

\bibitem{hastings} 
Hastings, W. K.: Biometrika, {\bf 57} (1970) 97.

\bibitem{sokal}
Sokal, A. D.: in ``Computer Simulation Studies in Condensed
Matter Physics: Recent Developments'', eds: D. P. Landau,
K. K. Mon, and H.-B. Sch\"uttler, 
Springer Proceedings in Physics, Vol {\bf 33}, (1988) 6.

\bibitem{onsager}
Onsager, L.: Phy. Rev. {\bf 65} (1944) 117.

\bibitem{swendsen-wang}
Swendsen, R. H., Wang, J.-S.: Phys. Rev. Lett. {\bf 58} (1987) 86.

\bibitem{wang-gan}
Wang, J.-S., Gan, C. K.: Phys. Rev. E. {\bf 57} (1998) 6548.

\bibitem{heermann} 
Heermann, D. W., Burkitt, A. N.: Physica A {\bf 162} (1990) 210.

\bibitem{coddington-PRL} 
Coddington, P. D., Bailie, C. F.: Phys. Rev. Lett. {\bf 68} (1992) 962.

\bibitem{kasteleyn}
Kasteleyn, P. W., Fortuin, C. M.: J. Phys. Soc. Jpn Suppl. {\bf 26} (1969) 11.

\bibitem{coniglio} Coniglio, A., Klein, W.: J. Phys. A {\bf 13} (1980) 2775.

\bibitem{hu} Hu, C.-K.: Phys. Rev. B {\bf 29} (1984) 5103, 5109.

\bibitem{edwards} 
Edwards, R. G., Sokal, A. D.: Phys. Rev. D {\bf 38} (1988) 2009.

\bibitem{wolff}
Wolff, U.: Phys. Rev. Lett. {\bf 62} (1989) 361.

\bibitem{tamayo}
Ray, T. S., Tamayo, P., Klein W.: Phys. Rev. A {\bf 39} (1989) 5949;
Tamayo, P., Brower R. C., Klein W.: J. Stat. Phys. {\bf 58} (1990) 1083.

\bibitem{wang-swendsen-kotecky} 
Wang, J.-S., Swendsen, R. H., Koteck\'y, R.: Phys. Rev. Lett. {\bf 63} 
(1989) 109; Phys. Rev. B {\bf 42} (1990) 2465.

\bibitem{brower}
Brower, R. C.: Phys. Rev. Lett. {\bf 62} (1989) 1087.

\bibitem{kandel}
Kandel, D., Ben-Av, R., Domany, E.: Phys. Rev. Lett. {\bf 65}
(1990) 941; Phys. Rev. B {\bf 45} (1992) 4700.

\bibitem{coddington}
Coddington, P. D., Han, L: Phys. Rev. B {\bf 50} (1994) 3058.

\bibitem{evertz}
Evertz, H. G., Luna, G., Marcu, M.: Phys. Rev. Lett. {\bf 70} (1993) 875.

\bibitem{dress}
Dress, D., Krauth, W.: J. Phys. A: Math. Gen. {\bf 28} (1995) L597.

\bibitem{wiese}
Wiese, U.-J., Ying, H.-P.: Z. Phys. B {\bf 93} (1994) 147.

\bibitem{kawashima-loop}
Kawashima, N., Gubernatis, J. E., Evertz, H. G.: Phys. Rev. B. {\bf 50} 
(1994) 136.

\bibitem{kawashima}
Kawashima, N., Gubernatis, J. E. Phys. Rev. E {\bf 51} (1995) 1547.

\bibitem{beard}
Beard, B. B., Wiese, U.-J.: Phys. Rev. Lett. {\bf 77} (1996) 5130.

\bibitem{creutz} 
Creutz, M.: Phys. Rev. Lett. {\bf 69} (1992) 1002.

\bibitem{heringa}
Heringa, J. R., Bl\"ote, H. W. J.: Phys. Rev. E {\bf 57} (1998) 4976;
   Physica A {\bf 254} (1998) 156.

\bibitem{bikker}
Bikker, R. P., Barkema, G. T.: Phys. Rev. E. {\bf 62} (2000) 5830.

\bibitem{machta}
Machta, J., Choi, Y. S., Lucke, A., Schweizer, T., Chayes, L. V.:
Phys. Rev. Lett. {\bf 75} (1995) 2792.

\bibitem{tomita}
Tomita, Y., Okabe, Y.: Phys. Rev. Lett. {\bf 86} (2001) 572.

\bibitem{besag}
Besag, J, Green, P. J.: J. R. Statis. Soc. B{\bf 55} (1993) 25, 53;
Gray, A. J.: Statistics and Computing, {\bf 4} (1994) 189;
Gaudron, I.: ESAIM: Probability and Statistics, {\bf 1} (1997) 259.

\bibitem{ferber}
von Ferber, C., W\"org\"otter, F.: Phys. Rev. E. {\bf 62} (2000) R1461.

\bibitem{gor}
Gore, V. K., Jerrum, M. R.: Proceeding of the 29th Annual ACM
Symposium on Theory of Computing (1997) 674; 
J. Stat. Phys. {\bf 97} (1999) 67.

\bibitem{sw-spin-glass}
Swendsen, R. H., Wang, J.-S.: Phys. Rev. Lett. {\bf 57} (1986) 2607.

\bibitem{liang}
Liang, S: Phys. Rev. Lett. {\bf 69} (1992) 2145.

\bibitem{random-field}
Machta, J., Newman, M. E. J., Chayes, L. B.:
Phys. Rev. E {\bf 62} (2000) 8782.

\bibitem{ferrenberg}
Ferrenberg, A. M., Swendsen, R. H.: 
Phys. Rev. Lett. {\bf 61} (1988) 2635.

\bibitem{fs-multiple}
Ferrenberg, A. M., Swendsen, R. H.:
Phys. Rev. Lett. {\bf 63} (1989) 1195 and 1658.

\bibitem{berg}
Berg, B. A., Neuhaus, T.: 
Phys. Rev. Lett.  {\bf 68} (1992) 9;  
Berg, B. A.: Inter. J. Mod. Phys. C {\bf 3} (1992) 311.

\bibitem{berg-JSP} 
Berg, B. A.: J. Stat. Phys. {\bf 82} (1996) 323.

\bibitem{berg-review}
Berg, B. A.: Fields Inst. Commun. {\bf 26}, 1 (2000).

\bibitem{lee} 
Lee, J.: Phys. Rev. Lett. {\bf 71} (1993) 211.

\bibitem{oliveira} 
de Oliveira, P. M. C., Penna, T. J. P.,
Herrmann, H. J.: Braz. J. Phys. {\bf 26} (1996) 677.

\bibitem{wang-euro}
Wang, J. S.: Eur. Phys. J. B,  {\bf 8} (1999) 287. 

\bibitem{wang-tay-swendsen-PRL} 
Wang, J.-S., Tay, T. K., Swendsen, R. H.:
Phys.  Rev. Lett. {\bf 82} (1999) 476; 
Wang, J.-S.: Comp. Phys. Commu. {\bf 121-122} (1999) 22.

\bibitem{oliveira-europhys-b-2} 
de Oliveira, P. M. C.:  Eur. Phys. J. B {\bf 6} (1998) 111.

\bibitem{berg-hansmann-europhys-b}  
Berg, B. A., Hansmann, U. H. E.: Eur. Phys. J. B {\bf 6} (1998) 395.

\bibitem{wang-lee} 
Wang, J.-S., Lee, L. W.: Comp. Phys. Commu. {\bf 127} (2000) 131;
Wang, J.-S.: Physica A {\bf 281} (2000) 174.

\bibitem{wang-landau} 
Wang, F., Landau, D. P.: Phys. Rev. Lett. (2001), to appear.

\bibitem{bortz} 
Bortz, A. B., Kalos, M. H., Lebowitz, J. L.:
J. Comput. Phys. {\bf 17} (1975) 10.

\bibitem{swendsen-int} 
Swendsen, R. H., Diggs, B., Wang, J.-S., Li, S.-T.,
Genovese, C., Kadane, J. B.:
Int. J. Mod. Phys. C {\bf 10} (1999) 1563. 

\bibitem{beale}
Ferdinand, A. E., Fisher, M. E.: Phys. Rev. {\bf 185} (1969) 832;
Beale, P. D.: Phys. Rev. Lett. {\bf 76} (1996) 78.

\bibitem{zhan} 
Zhan, Z. F., Lee, L. W., Wang, J.-S.:
Physica A {\bf 285} (2000) 239. 

\bibitem{munoz} 
Mu\~noz, J. D., Herrmann, H. J.: 
Int. J. Mod. Phys. C {\bf 10} (1999) 95;
Comput. Phys. Commu. {\bf 121-122} (1999) 13.

\bibitem{lima} 
Lima, A. R., de Oliveira, P. M. C., Penna, T. J. P.:
Sol. Stat. Commu. {\bf 114} (2000) 447.

\bibitem{kastner} 
Kastner, M., Promberger, M., Mu\~noz, J. D.:
Phys. Rev. E {\bf 62} (2000) 7422.

%
\end{thebibliography}
\end{document}